\documentclass[acmsmall,screen,nonacm]{acmart}

\usepackage{algorithmic,amsfonts,amsmath,amsthm,float,url, hyperref,comment}
\usepackage{cleveref,tikz}
\usetikzlibrary{shapes}
\crefformat{subsection}{\S#2#1#3}
\usepackage{listings}
\newtheorem{algorithm}{Algorithm}
\newtheorem{theorem}{Theorem}[section]
\newtheorem{lemma}[theorem]{Lemma}

\newtheorem{definition}[theorem]{Definition}

\newcommand{\RN}{\textnormal{RN}}

\newcommand{\ulp}{\textnormal{ulp}}
\newcommand{\lsb}{\textnormal{lsb}}
\newcommand{\F}{\mathbb{F}}

\floatstyle{ruled}
\newfloat{algorithm}{loa}{List of Algorithms}
\floatname{algorithm}{Algorithm}

\usepackage{color}
\definecolor{dkgreen}{rgb}{0,0.4,0}
\usepackage[utf8]{inputenc}
\usepackage{graphicx}
\usepackage{alltt}

\begin{document}
\title{Computing $ab+cd$ and other expressions with correct rounding in binary floating-point arithmetic}

\author{Stef Graillat}\affiliation{\institution{Sorbonne Université, CNRS, LIP6} \city{Paris}\country{France}}\orcid{0000-0001-8954-2276}\email{stef.graillat@lip6.fr}
\author{Jean-Michel Muller}\affiliation{\institution{CNRS, LIP, Université de Lyon} \city{Lyon}\country{France}}\orcid{0000-0003-3588-0047}\email{jean-michel.muller@cnrs.fr} 
\author{Takeshi Ogita}\affiliation{\institution{Waseda University}\city{Tokyo} \country{Japan}}\orcid{0000-0002-9346-2452}\email{t.ogita@waseda.jp}


\begin{abstract}
We introduce two ``high-level'' algorithms that make it possible to evaluate the sum $(x_h+x_\ell) + (y_h+y_\ell)$ of two double-word numbers with correct rounding in binary floating-point arithmetic. 
These algorithms allow one to evaluate expressions of the form $ab+cd$, $a+b+cd$, $a+b+c+d$, and $abc$  with correct rounding, where $a$, $b$, $c$, and $d$ are floating-point numbers.
\end{abstract}


\maketitle

\section{Introduction}

The expression $ab+cd$, where $a$, $b$, $c$, and $d$ are floating-point numbers, is ubiquitous in numerical computing: it appears in complex multiplication and division, Givens rotations, discriminants of quadratic equations, Fast Fourier Transforms, 2D rotations, etc. 
On current processors, that expression is almost never provided in hardware (an interesting exception is Kalray's MPPA3 processors, in binary32 arithmetic~\cite{dupontdedinechin:hal-04240798}). In this paper, we are interested in being able to evaluate this expression with \emph{correct rounding} in binary floating-point arithmetic (which means that we want to always return the floating-point number closest to the exact result). An interesting side effect of our study is that it can also be used to evaluate with correct rounding the expressions $a+b+cd$, $a+b+c+d$, and $abc$, which, although probably less ubiquitous than $ab+cd$, are also quite frequent in numerical computations. We are interested here in designing ``high-level'' algorithms, that is, algorithms that only use floating-point operations and comparisons, without manipulating the internal bit-string representations of the floating-point numbers. We believe that this property makes the corresponding algorithms more portable and easier to understand and debug.

\subsection{Notation and context}

In the following, we assume binary, precision-$p$ floating-point (FP) arithmetic, with round-to-nearest operations. We also assume an unbounded exponent range, which means that the results presented here apply to conventional binary floating-point arithmetic provided that underflow and overflow do not occur. In this context, a floating-point number $x$ is zero or a number of the form $M \times 2^{e_x-p+1}$, where $M, e_x \in \mathbb{Z}$ and $2^{p-1} \leq |M| \leq 2^p-1$. The \emph{FP exponent} of $x$ is $e_x$. We call $\F$ the set of the FP numbers.

We also assume that the rounding function is \emph{round-to-nearest, ties-to-even}, noted $\RN$, which is the default in IEEE 754 arithmetic~\cite{IEEE754-2019}. 

The  \emph{unit round-off} is $u = 2^{-p}$. It is an upper bound on the relative error due to rounding. We define the function $\ulp{}$ as
\[
\ulp(x) = \left\{
\begin{array}{ll}
0 & \textnormal{if } x=0, \\
2^{\left\lfloor \log_2 |x| \right\rfloor -p + 1} & \textnormal{otherwise}.
\end{array}
\right.
\]
If $x$ is a nonzero FP number, we also define the function $\lsb(x)$ as the largest power of $2$, say $2^k$, such that $x/2^k$ is an integer (i.e., $\lsb(x)$ is the weight of the least significant nonzero bit of $x$).
When an arithmetic operation $a \Delta b$ is performed (where $(a,b) \in \F^2$ and $\Delta \in \{+,-,\times,\div$\}), what is actually computed is $\RN(a \Delta b)$, and it satisfies
\[
\left|\RN(a \Delta b) - (a \Delta b)\right| \leq \frac{1}{2} \ulp(a \Delta b),
\]
and
\[
\left|\RN(a \Delta b) - (a \Delta b)\right| \leq \frac{u}{1+u} \cdot |a \Delta b| < u \cdot |a \Delta b|.
\]

We will say that the floating-point implementation $G$ of a real function $g$ is \emph{correctly rounded} if for any $x \in \mathbb{F}$, $G(x) = \RN(g(x))$.

A real number that is exactly halfway between two consecutive FP numbers is called a \emph{midpoint}. Function $\RN$ is discontinuous at midpoints, and constant in the open interval  between two consecutive midpoints: if there is no midpoint in the interval $[t_1,t_2]$ then $\RN(t_1)=\RN(t_2)$.

The \emph{FP predecessor} of $x \in \F$ is the largest FP number less than $x$. The \emph{FP successor} of $x$ is the smallest FP number larger than $x$.

We also assume that a correctly-rounded FMA (\emph{Fused Multiply-Add}) operator is available. This operator, when fed with $(a,b,c) \in \F^3$, returns $\RN(ab+c)$. The FMA instruction was introduced in 1990 in the IBM RS/6000 architecture~\cite{Mar90}. It has been specified by the IEEE 754 Standard on FP arithmetic since 2008, and is now available on most platforms (notable exceptions are the Java Virtual Machine and WebAssembly).

We will use the following classical algorithms (see~\cite{MullerEtAl2018} for an in-depth presentation and proofs).

\begin{algorithm}[H]
 \caption{\textbf{--~Fast2Sum$(a,b)$}. The Fast2Sum algorithm~\cite{Dekker71}.}
\label{alg:conventional-Fast2Sum}
\index{Fast2Sum}
  \begin{algorithmic}[0] 
    \STATE{$ s \gets \RN(a + b)$}
    \STATE{$ z \gets \RN(s - a)$}
    \STATE{$ t \gets \RN(b-z)$}
    \RETURN{$(s,t)$}
  \end{algorithmic}
\end{algorithm}

 If $(s,t) = \textnormal{Fast2Sum}(a,b)$ and if the
FP exponents $e_a$ and $e_b$ of $a$ and $b$ satisfy
$e_a \geq e_b$,  then $t$
is the error of the floating-point addition $\RN(a + b)$, that is, $a+b=s+t$.

\begin{algorithm}[H]
\caption{\textbf{--~2Sum$(a,b)$}. The 2Sum algorithm~\cite{Moller1965,Knu98}.}
\label{alg:twosum-conventional}
 \begin{algorithmic}[0]
   \STATE{$s \gets \RN(a + b)  $}
   \STATE{$a' \gets\RN(s - b)  $}
   \STATE{$b' \gets \RN(s - a')  $}
   \STATE{$\delta_a \gets \RN(a - a') $}
   \STATE{$\delta_b \gets \RN(b - b') $}
   \STATE{$t \gets \RN(\delta_a + \delta_b)$}
       \RETURN{$(s,t)$}
 \end{algorithmic}
\end{algorithm}

 For all FP numbers $a$ and $b$, if $(s,t) = \textnormal{2Sum}(a,b)$, then $t$ is the error of the floating-point addition $\RN(a + b)$.

\begin{algorithm}[H]
\caption{\textbf{--~2ProdFMA$(a,b)$}. The 2ProdFMA algorithm~\cite{Kah96}.}
\label{alg:2prodfma-conventional}
 \begin{algorithmic}[0]
   \STATE{$\pi \gets \RN(a \cdot b)  $}
    \STATE{$t \gets \RN(a \cdot b - \pi)$}
       \RETURN{$(\pi,t)$}
 \end{algorithmic}
\end{algorithm}

For all FP numbers $a$ and $b$, if $(\pi,t)= \textnormal{2ProdFMA}(a,b)$, then $t$ is the error of the floating-point multiplication $\RN(a \cdot b)$, that is, $a \cdot b=\pi+t$.

Finally, a pair $(x_h,x_\ell) \in \F^2$ is a \emph{Double Word} (DW) number if $x_h = \RN(x_h+x_\ell)$. The pairs $(s,t)$ returned by Algorithms~\ref{alg:conventional-Fast2Sum} and~\ref{alg:twosum-conventional}, and the pair $(\pi,t)$ returned by Algorithm~\ref{alg:2prodfma-conventional} are DW numbers. 

\subsection{Usual solutions for evaluating \texorpdfstring{$ab+cd$}{ab+cd}}

The naive solution for evaluating $ab+cd$ consists in computing
\begin{equation}
\label{naive-with-fma}
\RN(ab+\RN(cd)),
\end{equation}
if an FMA instruction is available, and
\begin{equation}
\label{naive-without-fma}
\RN\left(\RN(ab) + \RN(cd)\right),
\end{equation}
otherwise.

Both solutions are fairly accurate when $ab$ and $cd$ have the same sign. However, when $ab \approx -cd$, the results of (\ref{naive-with-fma}) and (\ref{naive-without-fma}) can be very poor in terms of relative error. A simple example in precision-$p$ floating-point arithmetic is $a = b = 2^{p-1}$, $c = -2^{p-1}+1$ and $d = 2^{p-1}+1$ for which $ab+cd = 1$, whereas we obtain $0$ when using (\ref{naive-with-fma}) or (\ref{naive-without-fma}). Various authors suggested ways of avoiding that kind of catastrophic loss of accuracy. Algorithm~\ref{alg:Kahan-abPluscd} below was attributed to Kahan by Higham~\cite[p. 60]{Hig02}.

\begin{algorithm}[H]
 \caption{\textbf{--~Kahan$(a,b,c,d)$}. Kahan's algorithm for $ab+cd$~\cite{Hig02}.}
\label{alg:Kahan-abPluscd}
  \begin{algorithmic}[0] 
    \STATE{$ (w,e) \gets \textnormal{2ProdFMA}(c,d)$}
    \STATE{$f \gets \RN(ab+w)$}
    \STATE{$x \gets \RN(f+e)$}
    \RETURN{$x$}
  \end{algorithmic}
\end{algorithm}

It was shown by Jeannerod et al.~\cite{JeannerodLouvetMullerMathComp2013} that the output $x$ of Algorithm~\ref{alg:Kahan-abPluscd} satisfies
\begin{equation}
\label{error-kahan}
\left\{
\begin{array}{lll}
\left| x - (ab+cd)\right| & \leq & 2u \cdot \left|ab+cd\right|, \\
\left| x - (ab+cd)\right| & \leq & \frac{3}{2} \ulp (ab+cd).
\end{array}
\right.
\end{equation}
This shows that Algorithm~\ref{alg:Kahan-abPluscd}  is much more accurate than the naive solution. The bound $\frac{3}{2} \ulp (ab+cd)$ is optimal: it is attained for some input values, such as $a=2^{p-1}+2^{p-3}$, $b = 2^{p-1}+1$, $c=2^{p-1}+2^{p-3}+2^{p-4}$, and $d = 2^{p-1}+2^{p-2}+1$. As a consequence, Algorithm~\ref{alg:Kahan-abPluscd} does not always return the correctly-rounded result $\RN(ab+cd)$ (since correct rounding would imply an error less than or equal to $\frac{1}{2}\ulp(ab+cd)$).

In their book \emph{Scientific Computing on Itanium-based Systems}~\cite{CorHarTan2002}, Cornea, Harrison, and Tang suggest another algorithm for evaluating $ab+cd$.
\begin{algorithm}[H]
 \caption{\textbf{--~CHT$(a,b,c,d)$}. Cornea, Harrison, and Tang's algorithm for $ab+cd$~\cite{CorHarTan2002}.}
\label{alg:CHT-abPluscd}
  \begin{algorithmic}[0] 
  \STATE{$\pi_1 \gets \RN(ab)$}
  \STATE{$e_1 \gets \RN(ab-\pi_1)$} // exact operation
  \STATE{$\pi_2 \gets \RN(cd)$}
   \STATE{$e_2 \gets \RN(cd-\pi_2)$} // exact operation
   \STATE{$\pi \gets \RN(\pi_1+\pi_2)$}
   \STATE{$e \gets \RN(e_1+e_2)$}
   \STATE{$x \gets \RN(\pi+e)$}
    \RETURN{$x$}
  \end{algorithmic}
\end{algorithm}  

Muller~\cite{Muller:2015:ECA:2732672.2629615} shows that the value $x$ returned by Algorithm~\ref{alg:CHT-abPluscd} satisfies
\begin{equation}
\label{error-CHT-1}
\left| x - (ab+cd)\right|  \leq  \left(2u+7u^2+6u^3\right) \cdot \left|ab+cd\right|,
\end{equation}
and Jeannerod~\cite{Jeannerod:2016:REA:2935754.2824252} slightly improves this bound to obtain
\begin{equation}
\label{error-CHT-2}
\left| x - (ab+cd)\right|  \leq 
\left\{
\begin{array}{ll} 
     2u \cdot \left|ab+cd\right| & \textnormal{ if } \RN{} \textnormal{ breaks ties to even,} \\
    \left(\frac{2u+u^2}{1-u^2}\right)  \cdot \left|ab+cd\right| & \textnormal{ otherwise. }
 \end{array}
 \right.
\end{equation}
Both relative error bounds~(\ref{error-CHT-1}) and~(\ref{error-CHT-2}) are  \emph{asymptotically optimal}. More precisely, with the input values $a=c=2^p-1$, $b=2^{p-3}+\frac{1}{2}$, and $d=2^{p-3}+\frac{1}{4}$, one has
\[
\left| x - (ab+cd)\right|  \leq \left(\frac{2u-3u^2}{1+2u-3u^2}\right) \cdot  \left|ab+cd\right| \approx \left(2u-7u^2\right) \cdot  \left|ab+cd\right|.
\]
This shows that Algorithm~\ref{alg:CHT-abPluscd} does not always return the correctly-rounded result $\RN(ab+cd)$.

Incidentally, since Algorithms~\ref{alg:Kahan-abPluscd} and~\ref{alg:CHT-abPluscd}  have approximately the same relative error bound, and since Algorithm~\ref{alg:CHT-abPluscd} requires more operations, one may ask what the point of that algorithm is. Its advantage over Algorithm~\ref{alg:Kahan-abPluscd} is that, since pairs $(a,b)$ and $(c,d)$ play a perfectly symmetrical role, it allows one to implement a complex multiplication that is commutative.

\section{Preliminary results}

\subsection{Reducing several problems to the evaluation of the sum of two DW numbers}

\begin{figure}[htb]\begin{footnotesize}
 \begin{tikzpicture}[scale=0.175, arr/.style={->,thick,shorten >=1pt,shorten <=1pt}]
  \node[circle,draw](abinfirstcorner) at (0,30){$ab$};
  \node (plusinfirstcorner) at (5,30){$+$};
  \node[circle,draw](cdinfirstcorner) at (10,30){$cd$};
  \node[ellipse,draw](DWx) at (20,20){$x_h+x_\ell$};
  \node(centralplus) at (30,20){$+$};
  \node[ellipse,draw](DWy) at (40,20){$y_h+y_\ell$};
  \draw[arr] (abinfirstcorner) to[out=-90,in=180] node[left,pos=0.3]{2ProdFMA} (DWx);
  \draw[arr] (cdinfirstcorner) to[out=0,in=120] node[above,pos=0.5]{2ProdFMA} (DWy);
  \node[ellipse,draw](aplusbinsecondcorner) at (48,30){$a+b$};
  \node (plusinsecondcorner) at (55,30){$+$};
  \node[ellipse,draw](cplusdinsecondcorner) at (62,30){$c+d$};
  \draw[arr] (aplusbinsecondcorner) to[out=-150,in=40] node[below right,pos=0.08]{2Sum} (DWx);
  \draw[arr] (cplusdinsecondcorner) to[out=-90,in=0] node[right,pos=0.25]{2Sum} (DWy);
  \node[ellipse,draw](aplusbinthirdcorner) at (0,10){$a+b$};
  \node (plusinthirdcorner) at (6,10){$+$};
  \node[circle,draw](cdinthirdcorner) at (10,10){$cd$};
  \draw[arr] (aplusbinthirdcorner) to[out=90,in=180] node[left,pos=0.3]{2Sum} (DWx);
  \draw[arr] (cdinthirdcorner) to[out=0,in=-120] node[below,pos=0.5]{2ProdFMA} (DWy);
  \node[circle,draw](abfourthcorner) at (43,0){$ab$};
  \node[circle,draw](cfourthcorner) at (48,0){$c$};
  \node(pihpluspil) at (43,10){$(\pi_h+\pi_\ell)$};
  \node(secondcfourthcorner) at (48,10){$c$};
  \draw[arr] (abfourthcorner)--(pihpluspil) node[left,midway]{2ProdFMA};
  \draw[arr] (cfourthcorner)--(secondcfourthcorner);
  \node(equal) at (51,10){$=$};
  \node[circle,draw](pihc) at (56,10){$\pi_hc$};
  \node(plusfourthcorner) at (60,10){$+$};
  \node[circle,draw](pilc) at (64,10){$\pi_\ell c$};
  \draw[arr] (pihc.north) .. controls (55,16) and (30,14.5) .. node[below,pos=0.45]{2ProdFMA} (DWx.south east);
  \draw[arr] (pilc.north) .. controls (64,18) and (52,17.5) .. node[right,pos=0.12]{2ProdFMA} (DWy.south east);
 \end{tikzpicture}
\end{footnotesize}
  \caption{Computing $\RN(ab+cd)$, $\RN(a+b+c+d)$, $\RN(a+b+cd)$, and $\RN(abc)$ reduce to computing $\RN\left((x_h+x_\ell) + (y_h+y_\ell)\right)$, where $(x_h,x_\ell)$ and $(y_h,y_\ell)$ are DW numbers.}
  \label{fig:reduction}
\end{figure}
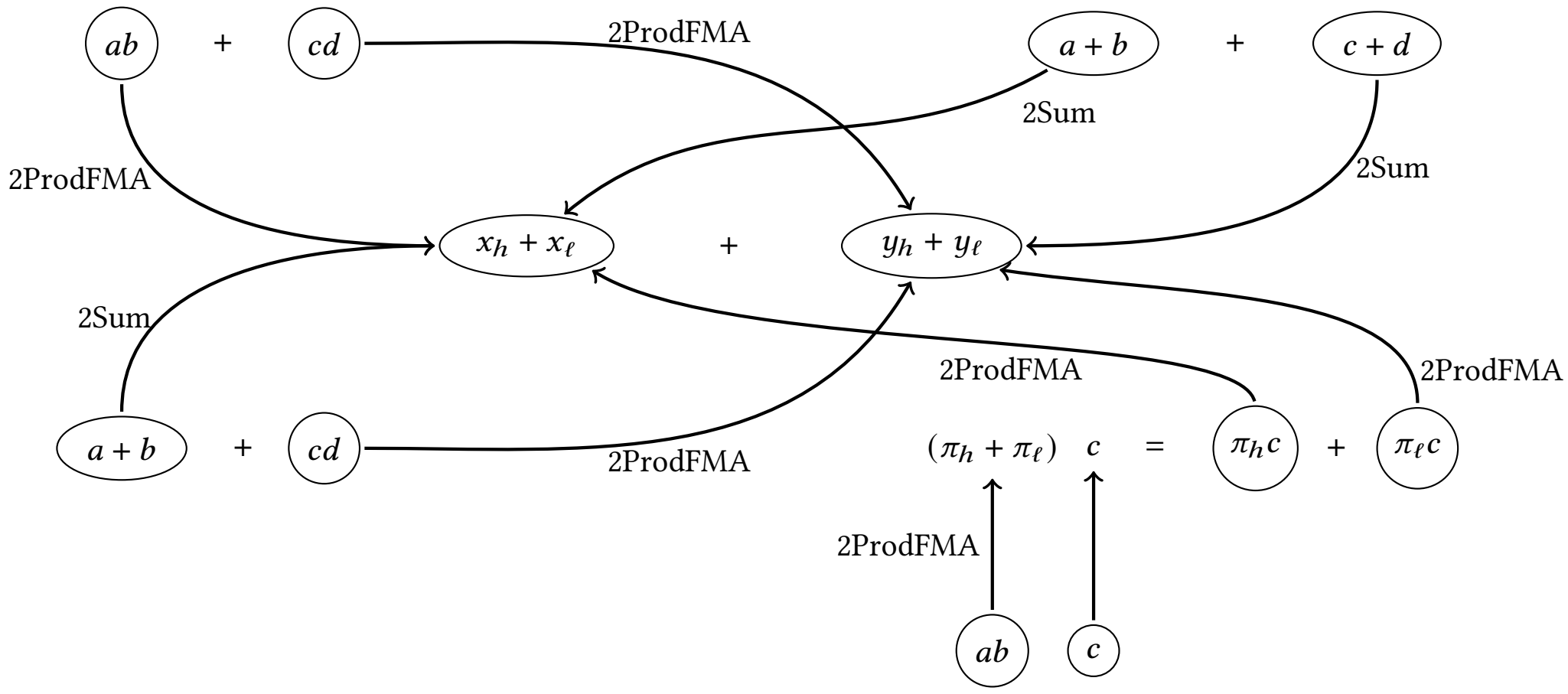

As Algorithms~\ref{alg:twosum-conventional} and \ref{alg:2prodfma-conventional} allow one to express the sum and product of two FP numbers exactly as a DW number, many calculations can be reduced to computing the correctly-rounded sum of two DW numbers: 
\begin{itemize}
    \item   for computing $\RN(ab+cd)$ one can transform the products $ab$ and $cd$ into DW numbers;
    \item for computing $\RN(a+b+c+d)$, it is the sums $a+b$ and $c+d$ that we can transform into DW numbers;
    \item the case of the computation of $\RN(a+b+cd)$ is straightforwardly similar;
    \item finally, to evaluate $\RN(abc)$, one can use Algorithm~\ref{alg:2prodfma-conventional} to express the product $ab$ as a DW, say $(\pi,t)$, and we are reduced to computing $\RN(\pi c + tc)$ (which is a particular case of $\RN(ab+cd)$).
\end{itemize} 
This is illustrated in Figure~\ref{fig:reduction}.

Hence, it suffices to focus on the evaluation of the correctly-rounded sum of two DW numbers, which will be done in Sections~\ref{sec:usingShewchuk} and~\ref{sec:firstalgorithm}.

  \subsection{Determining if a DW number is a midpoint}

  In the following, we will need to determine if a DW number is equal to a midpoint. If an FMA instruction is available, this can be done using the following result.

  \begin{theorem}
  \label{th:check-midpoint}
Assume $p \geq 2$. Let $a_h$ and $a_\ell$ be two FP numbers such that
\[
\RN(a_h+a_\ell) = a_h
\]
(i.e., the pair $(a_h,a_\ell)$ is a DW number).

Define
\[
\omega = \RN\left(a_\ell + (u+2u^2)a_\ell\right), \textnormal{ and}
\]
\[
a' = \RN\left(a_h + (1-u)\omega\right)
\]
(note that $\omega$ and $a'$ are easily computed with an FMA, since the constant, precomputed coefficients $u+2u^2$ and $1-u$ are FP numbers).

The number $a_h + a_\ell$ is a midpoint if and only if $a' \neq a_h$.
\end{theorem}

\begin{proof}
    Define
   \begin{equation}
   \label{eq:defchi}
    \chi = \left\{
    \begin{array}{ll}
     \frac{1}{4} \ulp(a_h) & \textnormal{if } a_h \textnormal{ is a power of } 2 \textnormal{ and } \\ 
       & a_h \textnormal{ and } a_\ell \textnormal{ have different signs,} \\ & \\
       \frac{1}{2} \ulp(a_h) & \textnormal{otherwise.}      
    \end{array}
    \right.
    \end{equation}
    \begin{itemize}
        \item If $a_h+a_\ell$ is a midpoint, then $|a_\ell| = \chi$. In that case (as $|a_\ell| = \chi$ is a power of $2$):
        \[
        \omega = \RN\left(a_\ell(1+u+2u^2)\right) = a_\ell \RN(1+u+2u^2) = a_\ell (1+2u),
        \]
        so that
        \[
        \left|(1-u) \omega\right| =  \left|(1-u)(1+2u) a_\ell\right| = (1+u-2u^2) \cdot \left|a_\ell\right|.
        \]
        Hence, as soon as $p \geq 2$ (i.e., $u \leq 1/4$), $\left|(1-u) \omega\right| > |a_\ell| = \chi$, so that $a' \neq a_h$.
        \item If $a_h+a_\ell$ is not a midpoint, then $|a_\ell| < \chi$. As $\chi$ is a power of $2$, the FP predecessor of $\chi$ is $\chi (1-u)$, so that we deduce:
        \[
        \left|a_\ell\right| \leq \chi (1-u).
        \]
        Hence $(1+u+2u^2) \left|a_\ell\right| \leq (1+u^2-2u^3) \chi,$ which implies
        \[
        |\omega| = \RN\left( (1+u+2u^2) \left|a_\ell\right|\right) \leq \RN \left((1+u^2-2u^3) \chi \right) = \RN \left(1+u^2-2u^3 \right) \cdot \chi = \chi,
        \]
        so that $|(1-u)\omega| < \chi$, which implies $a' = a_h$.
    \end{itemize}
\end{proof}

This gives the following algorithm.

\begin{algorithm}[H]
\caption{\textbf{--~IsMidpoint}$(e_h,e_\ell)$. \footnotesize{Checks if $e_h+e_\ell$ is a midpoint, where $e_h = \RN(e_h+e_\ell)$. It requires the precomputed constants $\phi_{1} = u + 2u^2$ and  $\phi_{2} = 1 - u$.}}
\label{alg:checkmidpoint4}
 \begin{algorithmic}[0]
\STATE{$\omega \gets \RN(e_\ell + \phi_{1}e_\ell)$} 
\STATE{$e' \gets \RN(e_h+ \phi_{2}\omega)$} 
\IF {$e' \neq e_h$} \RETURN{TRUE} \ELSE \RETURN{FALSE}
\ENDIF
\end{algorithmic}
\end{algorithm}

Algorithm~\ref{alg:checkmidpoint4} takes $2$ FP operations and $1$ test.

\subsection{A sufficient condition for the final rounding step}

In Sections~\ref{sec:usingShewchuk} and~\ref{sec:firstalgorithm}, we introduce two algorithms that compute the correctly rounded sum of two DW numbers. Both algorithms use the following lemma.

\begin{lemma}\label{lem:finalstep}
Assume $p \geq 2$. Let $(w_h,w_\ell)$ be a DW number with $w_\ell\neq 0$, let $r$ be a real number and let $v\in\mathbb{F}$ be such
that $v=0$ if and only if $r=0$, and $v$ and $r$ have the same sign. Assume that there exists a power of $2$, $\rho$,
such that $w_h$ and $w_\ell$ are multiples of $\rho$ and $|r|<\rho$. Define
\[
z = \begin{cases}
w_h & \text{if } \mathrm{IsMidpoint}(w_h,w_\ell) \text{ is false, or } v=0, \text{ or } (w_\ell>0) \mathbin{\mathrm{XOR}} (v>0),\\
\mathrm{RN}\!\left(w_h+\frac32 w_\ell\right) & \text{otherwise.}
\end{cases}
\]
Then $z=\mathrm{RN}(w_h+w_\ell+r)$.
\end{lemma}

\begin{proof}
Let $S = w_h+w_\ell+r$. Since $w_\ell\neq 0$, we have $w_h \neq 0$. If $r=0$, then $v=0$ and $z=w_h=\mathrm{RN}(w_h+w_\ell)=\mathrm{RN}(S)$,
which holds even when $w_h+w_\ell$ is a midpoint, since $w_h=\mathrm{RN}(w_h+w_\ell)$ already breaks the tie
to even. Assume now $r\neq 0$.
Let $m$ be the midpoint adjacent to $w_h$ on the side of $w_h+w_\ell$, and let $w'=w_h+2(m-w_h)$ be the FP neighbor of $w_h$ on
that side. Since $(w_h,w_\ell)$ is a DW number, $\rho\le|w_\ell|\le|m-w_h|$; as $|m-w_h|$ is a power of $2$, $m$ is a
multiple of $\rho$.
\begin{itemize}
\item If $w_h+w_\ell$ is not a midpoint, then $w_h+w_\ell$ lies strictly between $w_h$ and $m$, and since both $w_h+w_\ell$
and $m$ are multiples of $\rho$, $|m-(w_h+w_\ell)|\ge\rho>|r|$. As we also have $|r|<\rho\le|w_\ell|$, $S$ lies strictly
between $w_h$ and $m$, so that $\mathrm{RN}(S)=w_h$. By Theorem~2.1 (which requires $p \geq 2$), $\mathrm{IsMidpoint}(w_h,w_\ell)$ is false, hence $z=w_h$.
\item If $w_h+w_\ell=m$, then $w'=w_h+2w_\ell$ and, since $|r|<|w_\ell|$, $S=m+r$ lies strictly between $w_h$ and $w'$, an
interval whose only midpoint is $m$. Hence $\mathrm{RN}(S)=w'$ if $r$ (equivalently, $v$) has the sign of $w_\ell$, and
$\mathrm{RN}(S)=w_h$ otherwise. This is what is returned: $\mathrm{IsMidpoint}(w_h,w_\ell)$ is true, and
$w_h+\frac32w_\ell=m+\frac12w_\ell$ lies strictly between $m$ and $w'$, so that $\mathrm{RN}(w_h+\frac32w_\ell)=w'$.
\qedhere
\end{itemize}
\end{proof}

\section{First algorithm, derived from Shewchuk summation algorithm}
\label{sec:usingShewchuk}

In a seminal paper~\cite{She97}, Shewchuk presented a technique for adaptive precision arithmetic. One of the problems addressed in his paper was to somehow ``normalize'' a number expressed as the unevaluated sum of several floating-point numbers. To that end, he introduced a notion of ``nonoverlapping'' sequences of FP numbers that we recall in Definition~\ref{def:S-nonoverlapping} below. For more details, see \cite{She97,OgitaRumpOishi2005,BJMP17}.

Hereafter, we consider sequences of floating-point numbers $\{f_{1},f_{2},\dots,f_{n}\}$ whose components are sorted in order of decreasing magnitude\footnote{In Shewchuk's paper \cite{She97}, they are sorted in order of increasing magnitude.}, i.e.,
\[
  |f_{1}| \ge |f_{2}| \ge \dots \ge |f_{n}|
\]
with the important exception that \emph{any of the $f_{i}$ may be zero.}

\begin{figure}[t]
\begin{center}
\includegraphics[width=0.6\linewidth]{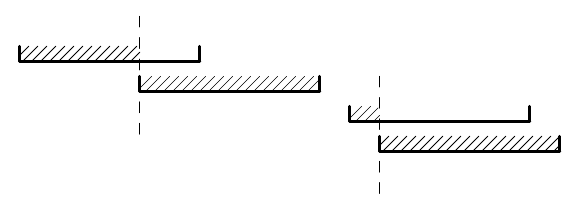}
\end{center}
\caption{S-nonoverlapping sequence by Shewchuk's scheme (cf.~\cite{OgitaRumpOishi2005})}
\label{fig:Shewchuk}
\end{figure}

\begin{definition}[S-nonoverlapping]
\label{def:S-nonoverlapping}
A sequence of floating-point numbers $\{f_{1},f_{2},\cdots,f_{n}\}$ is called \textit{S-nonoverlapping} if the magnitude of the least significant nonzero bit of $f_{i}$ for $1 \le i < n$ is greater than that of the most significant nonzero bit of $f_{i+j}$ for $1 \le j \le n - i$.
\end{definition}
An image of an S-nonoverlapping sequence is shown in Fig.~\ref{fig:Shewchuk}.

Note that if a pair $(x_1,x_2)$ is a DW number, then the sequence $\{x_1,x_2\}$ is S-nonoverlapping (but not vice versa), since $|x_2| \leq \frac{1}{2} \ulp(x_1) < \lsb(x_1)$.


Shewchuk proposed the algorithm \textbf{Expansion-Sum} \cite{She97} that adds two S-nonoverlapping sequences $e = \{e_{1},e_{2},\dots,e_{n}\}$ and $f = \{f_{1},f_{2},\dots,f_{m}\}$ and produces a new S-nonoverlapping sequence $h = \{h_{1},h_{2},\dots,h_{m+n}\}$ such that $\sum_{i = 1}^{m+n}h_{i} = \sum_{i = 1}^{n}e_{i} + \sum_{i = 1}^{m}f_{i}$  (see~\cite[Theorem~12]{She97}).

We now consider the special case $n = m = 2$, which transforms two S-nonoverlapping sequences $\{x_{h},x_{\ell}\}$ and $\{y_{h},y_{\ell}\}$ into a new S-nonoverlapping sequence $\{z_{1},z_{2},z_{3},z_{4}\}$ such that 
\[
  \sum_{i = 1}^{4}z_{i} = (x_{h} + x_{\ell}) + (y_{h} + y_{\ell}) .
\]

\begin{algorithm}[H]
\caption{\textbf{--~Expansion-Sum2}$(\{x_{h},x_{\ell}\},\{y_{h},y_{\ell}\})$. \footnotesize{Transforms two S-nonoverlapping sequences $\{x_{h},x_{\ell}\}$ and $\{y_{h},y_{\ell}\}$ into a new S-nonoverlapping sequence $\{z_{1},z_{2},z_{3},z_{4}\}$ such that $\sum_{i = 1}^{4}z_{i} = x_{h} + x_{\ell} + y_{h} + y_{\ell}$.}}
\label{alg:ExpansionSum2}
 \begin{algorithmic}[1]
    \STATE{$(\tau,z_{4}) \gets \textnormal{2Sum}(y_{\ell},x_{\ell})$} 
    \STATE{$(t_{h},t_{\ell}) \gets \textnormal{2Sum}(\tau,x_{h})$} 
    \STATE{$(\tau',z_{3}) \gets \textnormal{2Sum}(y_{h},t_{\ell})$} 
    \STATE{$(z_{1},z_{2}) \gets \textnormal{2Sum}(\tau',t_{h})$} 
   \RETURN{$\{z_{1},z_{2},z_{3},z_{4}\}$}
  \end{algorithmic} 
\end{algorithm}

Algorithm~\ref{alg:ExpansionSum2} takes $24$ FP operations.
Note that
\begin{itemize}
\item any of the $z_{i}$, $i = 1, 2, 3, 4$ may be zero,
\item the pair $(z_{1},z_{2})$ is a DW number, as it is the output of a $\textnormal{2Sum}$, and
\item the sequence $\{z_{1},z_{2},z_{3},z_{4}\}$ is S-nonoverlapping  by Theorem 12 in \cite{She97}.
\end{itemize}

Finally, we present the resulting algorithm for computing the correctly-rounded sum $\RN(x_h+x_\ell+y_h+y_\ell)$, where  $(x_h,x_\ell)$ and $(y_h,y_\ell)$ are two DW numbers.

\begin{algorithm}[H]
\caption{\textbf{--~CR-DWPlusDW-Shewchuk}$(x_h,x_\ell,y_h,y_\ell)$. \footnotesize{Computes $\mathrm{RN}(x_h+x_\ell+y_h+y_\ell)$, where $(x_h,x_\ell)$ and
$(y_h,y_\ell)$ are DW numbers.}}\label{alg:CR-DWPlusDW-Shewchuk}
\begin{algorithmic}[1]
\STATE $\{z_1,z_2,z_3,z_4\}\gets$ Expansion-Sum2$(\{x_h,x_\ell\},\{y_h,y_\ell\})$
\IF{$z_2=0$}
  \STATE $(w_h,w_\ell)\gets \mathrm{Fast2Sum}(z_1,z_3)$
  \STATE $v\gets z_4$
\ELSE
  \STATE $(w_h,w_\ell)\gets (z_1,z_2)$
  \STATE $v\gets \mathrm{RN}(z_3+z_4)$
\ENDIF
\IF{$w_\ell=0$}
  \STATE $s\gets\mathrm{RN}(w_h+v)$
\ELSIF{(\textbf{not} IsMidpoint$(w_h,w_\ell)$) \textbf{or} $v=0$ \textbf{or} $((w_\ell>0)\ \mathrm{\textbf{XOR}}\ (v>0))$}
  \STATE $s\gets w_h$
\ELSE
  \STATE $s\gets\mathrm{RN}(w_h+\frac32 w_\ell)$
\ENDIF
\RETURN $s$
\end{algorithmic}
\end{algorithm}

We have the following result.

\begin{theorem}\label{thm:alg9}
If $p \geq 2$, then for all pairs $(x_h,x_\ell)$ and $(y_h,y_\ell)$ of DW numbers, the floating-point number $s$ returned by
Algorithm~\ref{alg:CR-DWPlusDW-Shewchuk} is equal to $\mathrm{RN}(x_h+x_\ell+y_h+y_\ell)$.
\end{theorem}

\begin{proof}
Let $S=x_h+x_\ell+y_h+y_\ell=z_1+z_2+z_3+z_4$. We only use the fact that $\{z_1,z_2,z_3,z_4\}$ is S-nonoverlapping
(possible 21zero terms being ignored) and that $(z_1,z_2)$ is a DW number. If $f_i\neq0$, S-nonoverlapping implies that all nonzero
bits of the terms following $f_i$ have weights $<\lsb(f_i)$, and that these weights are pairwise distinct,
so that
\begin{equation}\label{eq:tail}
\Bigl|\sum_{j\ge1} f_{i+j}\Bigr| < \mathrm{lsb}(f_i),
\end{equation}
and every nonzero term preceding $f_i$ is a multiple of $\mathrm{lsb}(f_i)$.

\emph{Case $z_2\neq0$.} Then $z_1\neq 0$, $(w_h,w_\ell)=(z_1,z_2)$ is a DW number with $w_\ell\neq0$, and
$S=w_h+w_\ell+r$ with $r=z_3+z_4$. Let $\rho=\mathrm{lsb}(z_2)$: $z_1$ and $z_2$ are multiples of $\rho$ and, by
\eqref{eq:tail}, $|r|<\rho$. Since $v=\mathrm{RN}(r)$ has the sign of $r$ and vanishes iff $r=0$,
Lemma~\ref{lem:finalstep} gives $s=\mathrm{RN}(S)$.

\emph{Case $z_2=0$.} If $z_1=0$, Fast2Sum returns $(z_3,0)$ exactly. Otherwise, either $z_3=0$ or
$|z_3|<\mathrm{lsb}(z_1)\le|z_1|$, so that the exponent of $z_1$ is at least that of $z_3$ and Fast2Sum is exact.
In all cases $w_h+w_\ell=z_1+z_3$, $(w_h,w_\ell)$ is a DW number, and $S=w_h+w_\ell+z_4$ with $v=z_4$.
\begin{itemize}
\item If $w_\ell=0$, then $S=w_h+v$ and $s=\mathrm{RN}(w_h+v)=\mathrm{RN}(S)$.
\item If $w_\ell\neq0$, then $z_3\neq0$ (otherwise $z_1+z_3=z_1\in\mathbb{F}$). Let $\rho=\mathrm{lsb}(z_3)$: $z_1$ and
$z_3$ are multiples of $\rho$, hence so are $w_h=\mathrm{RN}(z_1+z_3)$ and $w_\ell$, and $|z_4|<\rho$ by
\eqref{eq:tail}. Lemma~\ref{lem:finalstep}, applied with $r=v=z_4$, gives $s=\mathrm{RN}(S)$.
\qedhere
\end{itemize}
\end{proof}

Algorithm~\ref{alg:CR-DWPlusDW-Shewchuk} takes at most $30$ FP operations and $5$ tests/comparisons.

In the following section, we present another, faster algorithm.


\section{Second algorithm}
\label{sec:firstalgorithm}

Let us now suggest another algorithm. As we are going to see, by adding the four terms $x_h$, $x_\ell$, $y_h$ and $y_\ell$ in a different order, we will be able to replace a 2Sum by a Fast2Sum and to simplify the necessary tests.

As before, we want to obtain $\RN(S)$, where $S = (x_h + x_\ell) + (y_h + y_\ell)$.
We suggest the following algorithm, that takes at most 25 FP operations and 4 tests.

\begin{algorithm}[H]
\caption{\textbf{--~CR-DWPlusDW}$(x_h,x_\ell,y_h,y_\ell)$. \footnotesize{Computes $\RN(x_h+x_\ell+y_h+y_\ell)$, where $(x_h,x_\ell)$ and $(y_h,y_\ell)$ are DW numbers.}}
\label{alg:dwplusdw}
 \begin{algorithmic}[0]
\STATE{$(s_h,s_\ell) \gets \textnormal{2Sum}(x_h,y_h)$}
\STATE{$(t_h,t_\ell) \gets \textnormal{2Sum}(x_\ell,y_\ell)$}
\STATE{$(e_h,e_\ell) \gets \textnormal{2Sum}(s_\ell,t_h)$}
\STATE{$(w_h,w_\ell) \gets \textnormal{Fast2Sum}(s_h,e_h)$}
\STATE{$v \gets \RN(e_\ell+t_\ell)$}
\IF{$w_\ell=0$}
\STATE{$z \gets \RN(w_h+v)$}
\ELSE
\IF{(\textbf{Not} IsMidpoint$(w_h,w_\ell)$)  \textbf{or} $v=0$ \textbf{or} (($w_\ell > 0$) \textbf{XOR} ($v  > 0$))}
\STATE{$z \gets w_h$}
\ELSE 
\STATE{$z \gets \RN(w_h + \frac{3}{2} w_\ell)$}
\ENDIF
\ENDIF
\RETURN{$z$}
 \end{algorithmic}
\end{algorithm}

Let us first explain why we can use the Fast2Sum algorithm for adding $s_h$ and $e_h$. 

\begin{lemma}
\label{lemma:Fast2SumOKinAlgDWPlusDW}
As soon as $p \geq 3$, the variables $s_h$ and $e_h$ in Algorithm~\ref{alg:dwplusdw} satisfy:
\begin{itemize}
   \item either $s_h = 0$;
   \item or $|e_h| \leq |s_h|$;
   \item or $e_h$ and $s_h$ have the same floating-point exponent.
 \end{itemize}
\end{lemma}

\begin{proof}
We assume $s_h \neq 0$. Let $\mu = \max\{\ulp(x_h),\ulp(y_h)\}$. 
\begin{enumerate}
   \item If $\ulp(x_h) = \ulp(y_h)$ then $x_h+y_h$ is a nonzero multiple of $\mu$, so that $s_h$ too is a nonzero multiple of $\mu$, and $|s_h| \geq \mu$. Note that $|s_h| = \mu$ is possible only when $|x_h+y_h| = \mu$, in which case $s_\ell = 0$ and therefore $e_h = t_h$. As $(x_h,x_\ell)$ and $(y_h,y_\ell)$ are DW numbers, we also have $|x_\ell| \leq \frac{\mu}{2}$ and $|y_\ell| \leq \frac{\mu}{2}$, and therefore $|t_h| \leq \mu$. We deduce:
   \begin{itemize}
       \item either $|s_h| = \mu$ and therefore $|e_h| = |t_h| \leq \mu$, so that $|e_h| \leq |s_h|$;
       \item or $|s_h| \geq 2\mu$. In that case, from $|t_h| \leq \mu \leq \left|\frac{s_h}{2}\right|$ and $|s_\ell| \leq u \cdot |s_h|$, we deduce $|t_h + s_\ell| \leq |s_h|$, so that $|e_h| \leq |s_h|$.
   \end{itemize}
   \item If $\ulp(x_h) \neq \ulp(y_h)$ then the smallest possible value of $|x_h+y_h|$ is $2^{p-1}\mu - (2^p-1) \frac{\mu}{2} =  \frac{\mu}{2}$. Therefore $|s_h| \geq \frac{\mu}{2}$. Also, as either ($|x_\ell| \leq \frac{\mu}{2}$ and $|y_\ell| \leq \frac{\mu}{4}$) or ($|y_\ell| \leq \frac{\mu}{2}$ and $|x_\ell| \leq \frac{\mu}{4}$), we have $|t_h| \leq \frac{3\mu}{4} \leq \frac{3}{2} |s_h|$. From $|s_\ell| \leq u |s_h|$, we finally obtain
   \[
   \left|t_h+s_\ell\right| \leq \left(\frac{3}{2}+u\right) |s_h|.
   \]
   Therefore,
   \begin{itemize}
       \item if  $\left|t_h+s_\ell\right| \leq |s_h|$ then $|e_h| \leq |s_h|$;
       \item if $|s_h| < |t_h+s_\ell|$ then $|e_h| = \RN(|t_h+s_\ell|) \geq \RN(|s_h|) = |s_h| \geq \mu/2$. Furthermore,
       \[
       \frac{\mu}{2} \leq |s_h| \leq |t_h+s_\ell| \leq \frac{3\mu}{4} + u \cdot |s_h|,
       \]
       in particular, this gives $|s_h|(1-u) \leq \frac{3\mu}{4}$, and therefore
       \[
       |t_h+s_\ell| \leq \frac{3\mu}{4}\left(1+\frac{u}{1-u}\right),
       \]
       which is $< \mu(1-u)$ as soon as $p \geq 3$.
       Hence, in that case, as $\mu$ is a power of $2$ (which implies that $\mu(1-u)$ is an FP number), $s_h$ and $|e_h| = |\RN(t_h+s_\ell)| \leq \mu(1-u)$ have the same FP exponent.
   \end{itemize}
\end{enumerate}
\end{proof}

We can now show that Algorithm~\ref{alg:dwplusdw} returns the desired result. 

\begin{theorem}
\label{th:A	lgDWplusDWiscorrect}
If $p \geq 6$, then for all pairs $(x_h,x_\ell)$ and $(y_h,y_\ell)$ of DW numbers,
the floating-point number $z$ returned by Algorithm~\ref{alg:dwplusdw}  is equal to $\RN(x_h+x_\ell+y_h+y_\ell)$.
\end{theorem}

\begin{proof}
We use the following classical facts (no underflow): $\mathrm{RN}$ is monotone and odd,
$\mathrm{RN}(x)=0 \Leftrightarrow x=0$, $|\mathrm{RN}(x)-x| \le \frac{u}{1+u}|x|$,
$\mathrm{ulp}(x) > u|x|$ for $x \neq 0$; if $x$ is a multiple of a power of $2$, say $\rho$, then so is
$\mathrm{RN}(x)$; and if $(a_h,a_\ell)$ is a DW number then $|a_\ell| \le \frac{1}{2}\mathrm{ulp}(a_h)$.

By Lemma~\ref{lemma:Fast2SumOKinAlgDWPlusDW}, Fast2Sum is exact, so $w_h+w_\ell = s_h+e_h$ and $(w_h,w_\ell)$ is a DW number. Hence
\begin{equation}\label{eq:S}
S = x_h+x_\ell+y_h+y_\ell = w_h + w_\ell + r, \quad\text{with } r = e_\ell + t_\ell \text{ and } v=\mathrm{RN}(r).
\end{equation}
Note that $v$ and $r$ have the same sign, and that $v=0$ iff $r=0$.

\emph{Step 1: bounds on $r$.}

(a) Assume $t_\ell \neq 0$ (otherwise what follows is trivial). Then $x_\ell, y_\ell, t_h, x_h, y_h$ are nonzero.
Let $\lambda = \min\{\mathrm{ulp}(x_h),\mathrm{ulp}(y_h)\}$ and $\eta=\min\{\lambda,\mathrm{ulp}(t_h)\}$.
Since $x_\ell$ and $y_\ell$ are FP numbers and $t_h = \mathrm{RN}(x_\ell+y_\ell)$, we have
$|t_\ell| \le \min\{|x_\ell|,|y_\ell|\} \le \lambda/2$, and $|t_\ell| \le \frac{1}{2}\mathrm{ulp}(t_h)$, so $|t_\ell|\le \eta/2$.
The numbers $s_h$, $s_\ell = x_h+y_h-s_h$ and $t_h$ are multiples of $\eta$, hence so are $e_h$, $e_\ell$, $w_h$ and $w_\ell$.
Consequently: if $e_\ell \neq 0$ then $|t_\ell| \le |e_\ell|/2$; if $e_\ell = 0$ then $|r| = |t_\ell| \le \eta/2 < \eta$.

(b) Assume $e_\ell \neq 0$. Then $s_\ell \neq 0$ (otherwise $e_h = t_h$ exactly), so $s_h \neq 0$, $e_h \neq 0$ and
$x_h + y_h \notin \mathbb{F}$. Let $\xi = \min\{\mathrm{ulp}(s_h),\mathrm{ulp}(e_h)\}$. Since $s_h$ and $e_h$ are multiples of $\xi$, $w_h$ and $w_\ell$ are also multiples of $\xi$.
Since $t_h\in\mathbb{F}$, $|e_\ell| \le |s_\ell| \le \frac12 \mathrm{ulp}(s_h)$, and $|e_\ell| \le \frac12\mathrm{ulp}(e_h)$, so
$|e_\ell| \le \xi/2$ and, by (a), $|r| \le \frac{3}{2}|e_\ell| \le \frac{3}{4}\xi < \xi$.

Let us also bound $r$ relative to $s_h$. Assume, without loss of generality, $\mu = \mathrm{ulp}(x_h) \ge \mathrm{ulp}(y_h)$.
If $\mathrm{ulp}(y_h)=\mu$, then $x_h+y_h$ is a multiple of $\mu$ that is not in $\mathbb{F}$, so $|x_h+y_h| > 2^p\mu$.
Otherwise $|y_h| < |x_h|$; by Sterbenz's lemma we cannot have $|y_h|\ge |x_h|/2$ with $x_hy_h<0$, so
$|x_h+y_h| \ge |x_h|/2 \ge 2^{p-2}\mu$. In all cases $|s_h| \ge 2^{p-2}\mu$, i.e., $\mu \le 4u|s_h|$.
As $|x_\ell|,|y_\ell| \le \mu/2$ and $\mu \in \mathbb{F}$, $|t_h| \le \mu$, so that
$|s_\ell + t_h| \le u|s_h| + \mu \le 5u|s_h|$. Therefore $|e_\ell| \le 5u^2|s_h|$, $|e_h| \le 5u(1+u)|s_h|$, and
\[
|r| \le \tfrac{15}{2}u^2|s_h|, \qquad |v| \le (1+u)|r| \le 8u^2|s_h|, \qquad
|w_h| \ge \frac{|s_h+e_h|}{1+u} \ge \frac{1-5u-5u^2}{1+u}\,|s_h|.
\]
For $p\ge 6$ we have $37u+37u^2 \le 1$, i.e., $8u^2 \le \frac{u}{4}\cdot\frac{1-5u-5u^2}{1+u}$, so
$\max\{|r|,|v|\} \le \frac{u}{4}|w_h| < \frac14\mathrm{ulp}(w_h)$. Since the midpoints closest to $w_h$ are at distance
at least $\frac14\mathrm{ulp}(w_h)$ from it, we obtain
\begin{equation}\label{eq:abs}
e_\ell \neq 0 \;\Longrightarrow\; \mathrm{RN}(w_h + r) = \mathrm{RN}(w_h+v) = w_h.
\end{equation}

\emph{Step 2: case $w_\ell = 0$.} The algorithm returns $z=\mathrm{RN}(w_h+v)$ and $S = w_h + r$.
If $e_\ell=0$, then $v = \mathrm{RN}(t_\ell) = t_\ell = r$, so $z = \mathrm{RN}(S)$. If $e_\ell \neq 0$, \eqref{eq:abs}
gives $z = w_h = \mathrm{RN}(S)$.

 \emph{Step 3: case $w_\ell\neq 0$.} If $r = 0$, take $\rho = \mathrm{lsb}(w_\ell)$; otherwise, by Step~1, $w_h$ and $w_\ell$ are
 multiples of $\rho=\eta$ (if $e_\ell=0$) or $\rho=\xi$ (if $e_\ell\neq0$), and $|r|<\rho$. Since $v=\mathrm{RN}(r)$
 has the sign of $r$ and vanishes iff $r=0$, Lemma~\ref{lem:finalstep} gives $z=\mathrm{RN}(S)$.

\end{proof}

\section{Experiments}

We implemented the computation of $\RN(ab+cd)$ in binary64 arithmetic using Algorithms~\ref{alg:CR-DWPlusDW-Shewchuk} and~\ref{alg:dwplusdw} in C. To check the correctness of the algorithms and their implementation, we compared their results with the results returned by the multiple-precision GNU MPFR software on $10^7$ randomly-chosen inputs, and on $10^7$ specifically built ``difficult'' inputs (for which $ab \approx -cd$). The two algorithms and MPFR agreed on all input values.

Concerning timings, we compared the times needed to perform $5 \times 10^9$ calculations $ab+cd$ in binary64 arithmetic, using  Algorithm~\ref{alg:CR-DWPlusDW-Shewchuk}, Algorithm~\ref{alg:dwplusdw}, and   Kahan's Algorithm (Algorithm~\ref{alg:Kahan-abPluscd}, which is not correctly rounded), on different computer architectures. The compilation options were \texttt{-O3 -ffp-contract=off -march=native}. As there are no branches in Kahan's algorithm, it can be very efficiently vectorized, which is not the case for our algorithms. To be able to compare the algorithms without vectorization, we also measured the performance of Kahan's algorithm with the compilation option \texttt{-fno-tree-vectorize}. The resulting timings are reported in Table~\ref{tabcompar-errfma}. As suggested by the number of FP operations required, Algorithm~\ref{alg:dwplusdw} is significantly faster (between $23\%$ and $40\%$ faster) than Algorithm~\ref{alg:CR-DWPlusDW-Shewchuk}. Even without vectorization, Kahan's algorithm is much faster than our algorithms (about $4.4$ to $5.2$ times faster than Algorithm~\ref{alg:dwplusdw}), but it does not always deliver a correctly rounded result. To give an idea of how often Kahan’s algorithm fails to return a correctly rounded result, we compared it with MPFR on $10^7$ inputs for several distributions. With the distribution of Table 1, the result is incorrectly rounded in about $4.9\%$ of the cases; when all operands are uniformly distributed in $\pm[1,2)$, this percentage rises to about $18.7\%$. The largest observed error, $1.5~\ulp$, matches the optimal bound (\ref{error-kahan}).

%

\begin{table}[H]
\caption{\footnotesize{Time (in seconds) to perform $5 \times 10^9$ calculations $ab+cd$ in binary64 arithmetic, using  Algorithm~\ref{alg:CR-DWPlusDW-Shewchuk}, Algorithm~\ref{alg:dwplusdw}, and  Kahan's Algorithm (Algorithm~\ref{alg:Kahan-abPluscd}, which is not correctly rounded), on different computer architectures and with different compilers. Each operand is a random \texttt{double} of the form $K \times s \times F$, where $F$ is uniform in $[0,1]$, $s$ is $-1$ or $+1$ (each with probability $1/2$), and $K \in \{1, 2^{20}, 2^{-20}, 2^{40}, 2^{-40}, 2^{60}, 2^{-60}, 2^{80}, 2^{-80}\}$ (each with probability $1/9$).} We also measured the performance of these algorithms on ``difficult'' cases ($ab \approx -cd$) and obtained similar figures.}
\label{tabcompar-errfma}
\begin{footnotesize}
\begin{center}
\begin{tabular}{|l|l|l|l|l|l|}
\hline
Architecture/System & compiler & Algorithm~\ref{alg:CR-DWPlusDW-Shewchuk} & Algorithm~\ref{alg:dwplusdw}  & Kahan's algorithm  &  Kahan's algorithm\\
& & (CR) & (CR) & (not CR) & with vectorization disabled \\
  \hline
 Apple M3Pro  (MacOS)& clang  (v. 17.0.0) &  11.34   &8.93  & 1.08 & 1.85\\
  \hline
 AMD Opteron 6272 (Linux) & gcc (v. 12.2.0)&  98.0 & 69.9 &8.92 & 15.91 \\
 \hline
 Intel Xeon  Gold 6444Y (Linux)  & gcc  (v. 14.2.0)& 20.35 &16.54 &  0.81 & 3.20\\
 \hline
\end{tabular}
\end{center}
\end{footnotesize}
\end{table}
 \section*{Conclusion}

We have presented two algorithms for computing the correctly-rounded sum of two DW numbers. These algorithms allow the evaluation in floating-point arithmetic of expressions of the form  $ab+cd$, $a+b+cd$, $a+b+c+d$, and $abc$ with correct rounding. Our tests show that these algorithms are reasonably efficient. In particular, they may be used to implement very accurate algorithms for complex-number arithmetic.

\section*{Note}
We received  help from Claude, Anthropic’s AI assistant, for checking the preliminary (and in rare cases, wrong!) versions of Algorithms~\ref{alg:CR-DWPlusDW-Shewchuk} and~\ref{alg:dwplusdw}, exploring variants, checking and improving the proofs, and designing the C code for the experiments.

 \section*{Acknowledgement}
 This work was partly supported by the NuSCAP (ANR-20-CE48-0014) project, the InterFLOP (ANR-20-CE46-0009) project and the FPT-4 project (ANR-24-CE46-7572) of the French National Agency for Research (ANR).

\bibliographystyle{plain}
\bibliography{biblio}

\end{document}